\documentclass{edbk} 
\usepackage[dvips]{graphicx}

\normallatexbib

\begin{document}


\articletitle[]{Imagery of diffusing media \\
by heterodyne holography}

\author{Michel Gross, Fr\'ed\'erique Le Clerc }
\affil{Laboratoire Kastler Brossel de L'Ecole Normale Supérieure \\
CNRS- UMR 8552 -Univertit\'e-CNRS  Paris 6, \\
24 rue Lhomond 75231 Paris Cedex 05, France }
\email{gross@lkb.ens.fr, leclerc@lkb.ens.fr}%

\author{Laurent Collot}
\affil{Thomson CSF Optronique, \\
Rue Guynemer BP 55, 78 283 Guyancourt, France}
\email{lcollot@club-internet.fr}%


\chaptitlerunninghead{Heterodyne Holography Imaging}





\section{HETERODYNE HOLOGRAPHY }

\begin{figure}
\begin{center}
\includegraphics[width=7cm,keepaspectratio=true]{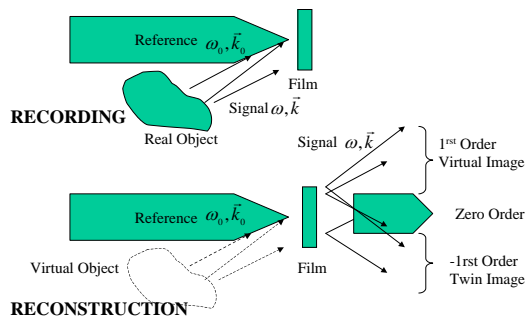}
\caption{Thin film off-axis Holography: $ \omega  = \omega _0 $, $
k \ne k_0 $}\label{gr_homo}
\end{center}
\end{figure}

Demonstrated by Gabor \cite{Gabor49} in the early 50's, the
purpose of holography\inxx{holography} is to record on a 2D
detector the phase and the amplitude of the light coming from an
object under coherent illumination. Thin film holography (see
Fig.\ref{gr_homo}) does not provide a direct access to the
recorded data. Numerical holography\inxx{holography, numerical}
\cite{Macovsky1971} replaces the holographic film by a 2D
electronic detector allowing quantitative numerical analysis.
This simple idea needed the recent development of computer and
video technology to be experimentally
demonstrated\inxx{holography,digital} \cite{Schnars94}. In both
thin film and numerical holography the system records only one
interference phase state, that allows to calculate only one
quadrature of the complex field. This incomplete measurement
yields to a  virtual image ($order=1$) of the object that is
superposed \cite{Kreis88} with a ghost twin image ($order=-1$)
and with the  remaining part of the reference field ($order=0$).
A solution to this problem (see Fig.\ref{gr_homo}) is to tilt the
reference beam in respect to the signal beam \cite{Leith62} in
order to separate physically the three images and to select the
wanted image. This off-axis
method\inxx{holography,digital,off-axis} reduces the useful
angular field of view and restricts measurement to the far field
region where the 3 images are spatially separated.

In order to avoid these problems it is necessary to get more
information by recording more than one phase state of the
interference holographic pattern . A possible method is to record
several holograms while shifting the phase of the reference beam
with a PZT mirror\inxx{holography, digital, on-axis}
\cite{Yamaguchi1997}. We have developed an alternate technique
that we call Heterodyne Holography\inxx{holography,heterodyne}
\cite{LeClerc2000} where we record on a CCD camera the
interference of the signal field with a reference field, which is
frequency shifted by $ \delta f $. Each pixel of the CCD camera
performs thus heterodyne detection of the signal field. To make
the demodulation procedure easier the heterodyne frequency $
\delta f = 2\pi (\omega _0  - \omega ) = 6.25Hz $ is chosen equal
to $ 1/4 $ of the $ 25 Hz $ video frame rate. If the reference is
a plane wave, the complex signal field $ E_s $ is proportional to
$ (I0 - I2) + i.(I3 - I4)$ where $I0$, $I1$, $I2$, $I3$ are four
successive video images.

\begin{figure}
\begin{center}
\includegraphics[height=5cm,keepaspectratio=true]{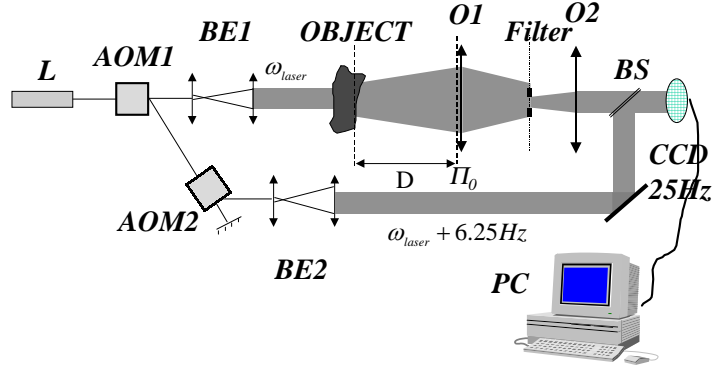}
\caption{Heterodyne Holography Setup: L HeNe laser, AOM1 and AOM2 acousto-optic modulator, BE1 and BE2 %
beam expanders, O1 and O2 confocal objectives, BS beam splitter %
and CCD detector.}\label{setup}
\end{center}
\end{figure}

Our heterodyne holography experimental setup is shown in Fig.
\ref{setup}. Although we consider here a transmission setup, most
of the following remain valid in a reflection configuration. The
coherent source is an Helium Neon laser L. The two accousto-optic
modulators AOM1 and AOM2, which are working at $80 MHz$ and
$80MHz + 6.25Hz$ respectively, provide the frequency shift of the
reference beam. The reference and signal beam are expanded by 2
beam expanders BE1 and BE2. The 2 beams are combined by the beam
splitter BS and the interference pattern is recorded by the CCD
camera. A frame grabber transfers the information to the computer
PC. Since the CCD pixel spacing is finite (typically $10\mu m$),
the CCD camera performs a spatial sampling of the field. The
sampling theorem limits thus the field of view angle $\theta$ for
valid measurements to:

\begin{equation}\label{sampling}
\left| \theta  \right| \le \theta _{\max }  = {\lambda
\mathord{\left/
 {\vphantom {\lambda  {(2d_{pixel} )}}} \right.
 \kern-\nulldelimiterspace} {(2d_{pixel} )}}
\end{equation}

where $d_{pixel}$ is the CCD pixel spacing. One can notice that
this sampling condition is common to both on and off-axis digital
holography. In order to fulfill Eq.\ref{sampling} we have
selected, in our experimental setup, the near axis photons by a
spatial filter system ($O1$, $O2$, $Filter$ on Fig.\ref{setup}).

Detailed tests and discussions on our system for holography are
given in \cite{LeClerc2000}. We show that heterodyne holography
performs within the spatial filter selected region, a complete
measurement of the signal field without information loss. Since
the technique is sensitive to the field amplitude (heterodyne
detection), one photon detection is possible. The dynamic range is
limited by the number $ n \approx 3.10^5 $ of electrons that can
be stored on each CCD pixel, and corresponds for each pixel to
about $ {n \mathord{\left/
 {\vphantom {n {\sqrt n }}} \right.
 \kern-\nulldelimiterspace} {\sqrt n }} \approx 5.10^2
$, i.e to $9$ bit data, or to $54$ $dB$. Another important point
is information. Our system grabs in one second $12. 10^6$ words
with $16$ bit, and extracts $6.25$ complex field images with $5.
10^5$ pixels.

\section{APPLICATION TO DIFFUSING MEDIA }

In tissues, diffusion of the light is mainly related to the small
( $ \approx 5$  to  $10\%$ ) change of the refractive index
within each cell. As these changes occur over distances larger
than the wavelength, scattering is highly directive
\cite{vandeHulst1957} in the forward direction and the 2 lengths
$ l_s $ and $ l'_s $ that govern scattering are very different .
$ l_s \simeq 50 $ $ \mu m$ is the scattering length, i.e. the
length beyond which the light phase is lost, and $ l'_s \simeq 1
$ $ mm $ is the  light transport mean free path, i.e. the length
for loosing the propagation direction.

In this context, heterodyne detection has been used to select the
photons transmitted through a diffusing media, which remains
coherent. Using this technique Inaba\inxx{heterodyne,detection}
\cite{inaba} gets quite nice images when scanning a mono pixel
detector (photodiode). Here we go further and perform heterodyne
detection on 2D\inxx{heterodyne,detection,2D} detectors by using
our heterodyne holography technique. Contrarily to Inaba we record
holograms where the pixel to pixel relative phase remains
meaningful. Our idea is to acquire a maximum amount of
information on a diffusing object, in order to extract later
useful pertinent results.

\begin{figure}
\begin{center}
\includegraphics[width=10 cm,keepaspectratio=true]{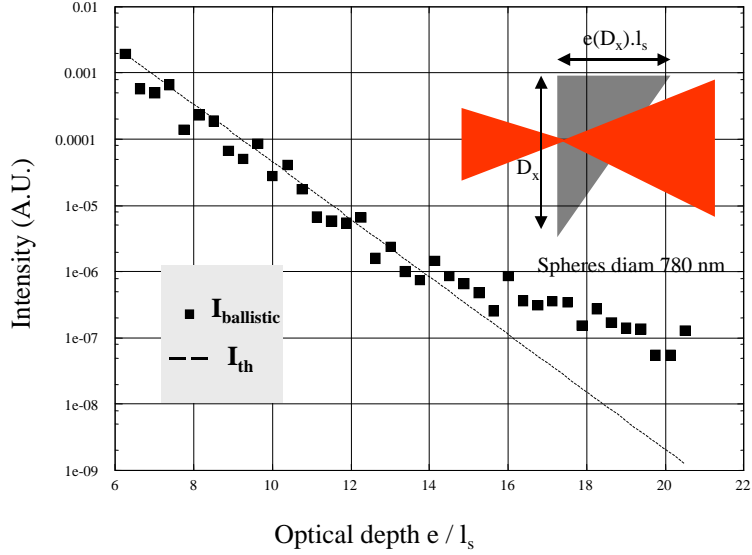}
\caption{Ballistic photons in gel.  }\label{gel}
\end{center}
\end{figure}

\begin{figure}
\begin{center}
\includegraphics[width=10 cm,keepaspectratio=true]{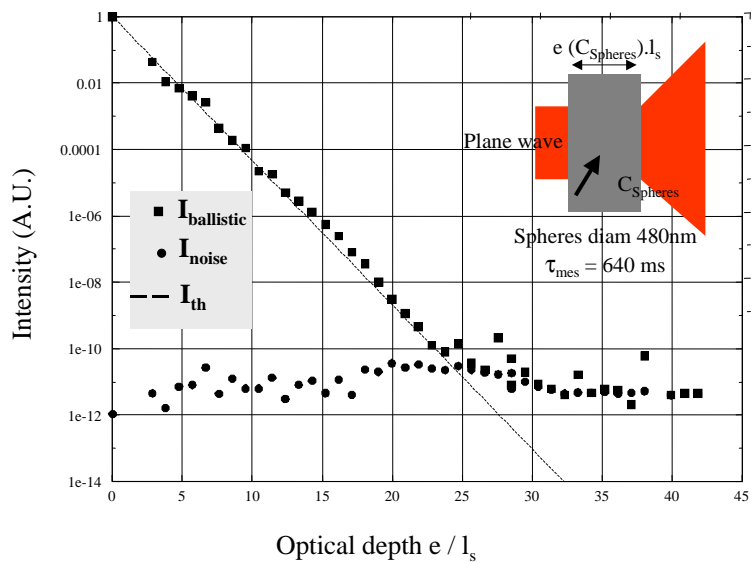}
\caption{Ballistic photons in liquid. }\label{liquid}
\end{center}
\end{figure}

To illustrate this, we have performed an heterodyne holography
experiment on latex sphere\inxx{latex sphere} solutions in
gelinxx{latex sphere,in gel} and liquid{latex sphere,in liquid},
and we have selected in the transmitted signal field, the
ballistic\inxx{ballistic, photon} component that correspond to the
photons, which have passed through the solution without
interacting. In gel, the concentration of the $780$ $nm$ sphere
solution is kept constant while the light is focused on the
surface of the prism shaped solution cell. Translating the cell
in the transverse direction changes the medium effective depth
(see Fig. \ref{gel}). In liquid the cell that contains the $480$
$nm$ sphere solution is rectangular. The cell is illuminated by
plane light beam and the effective depth of the medium is altered
by changing the concentration of the solution (see Fig.
\ref{liquid} ).

The diffusing medium transmitted field is first measured. The
weight of the ballistic photons components is then determined by
calculating the correlation of the measured field, which is the
sum of a ballistic and a diffused component, with the pure
ballistic reference field, that corresponds to the field measured
without scatters. In Fig.\ref{gel} and \ref{liquid}  the relative
(with respect to the incident field) weight of the ballistic
component (black squares) is plotted in log scale (y-axis), as a
function of the depth of the medium in scattering length unit $
l_s $ (x-axis). As expected, for both liquid and gel, the
ballistic component decreases exponentially with a length
parameter $ l_s $ (dotted line in Fig.\ref{gel} and
Fig.\ref{liquid}). The noise floor is about $70$ $dB$ ($ 10^{
-7}$) and $110$ $dB$ ($10^{ -11}$) below the incident field for
gel and liquid respectively. This correspond to an effective depth
of about $16$ $ l_s $ and $ 25$ $ l_s$ respectively. These results
can be compared favorably with \cite{Genack1997}, who perform the
equivalent experiment with pinhole selection of the ballistic
photons.

Our results can be understood quite easily. As the diffusing depth
is large, most of the incident photons are back reflected, and the
small transmitted component is spread over a large solid angle. A
small part $p$ of the incident photon ($p \approx 10^{ -2}$ to
$10^{ -3}$) passes thus through the Fig.\ref{setup} spatial
filter (that corresponds to Eq.\ref{sampling} sampling
condition), reaches the CCD, and is detected. By Fourier
transform, we have calculated the decomposition of the detected
signal over the k-modes. The diffused component is randomly
spread with equal weight over all the k-modes, while the
ballistic component remains in the k-mode that correspond to the
incident field. As the number of k-modes N is simply equal to the
number of pixels ($5.10^ { 5}$), the noise floor, which
corresponds to the weight of the diffused component within the
ballistic k-mode (or any k-mode), is expected to be simply $ p/N
\approx 2.10^{ - 7}$ to $2.10^{ - 8}$. This result is in good
agreement with the gel experiment.

In liquids, the experiment overcomes this limit, because the
Brownian motion\inxx{Brownian, motion} of the scatters shifts the
frequency of the scattered light. As our heterodyne system has an
extremely narrow detection bandwidth (that is equal to the
inverse of the measurement time: $640$ $ms$), most of the
diffused photons are not detected. We have plotted (black circle
on Fig.\ref{liquid}), for liquid, the average value of the noise
per mode. As expected, the ballistic noise floor corresponds to
this one mode noise.

\begin{figure}
\begin{center}
\includegraphics[width=9 cm,keepaspectratio=true]{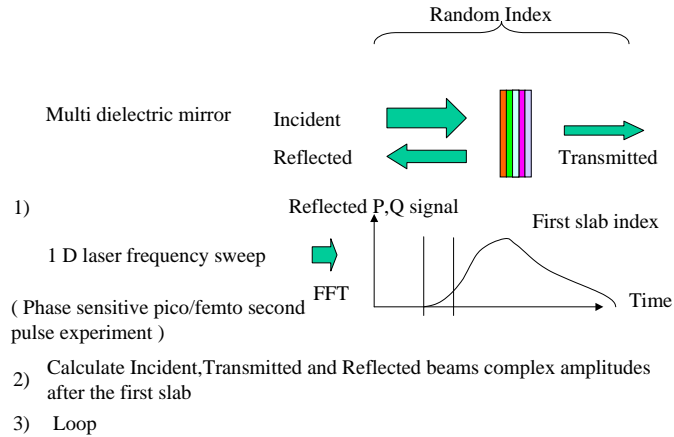}
\caption{Solving the 1D diffusion inverse problem) }\label{diff1d}
\end{center}
\end{figure}

Heterodyne holography may also be used in a more ambitious way.
The idea is to perform many measurements on a diffusing object in
order to solve the inverse diffusion problem\inxx{inverse
problem,1D}. Let's discuss this point on the simpler 1D
"gedanken" experiment . Let's perform heterodyne holography on a
1D diffusing object as, for example, a random index flat
multi-layer mirror (see Fig.\ref{diff1d}). The 2D detector
becomes here a zero D mono pixel detector. To calculate the
refractive index for the M slabs of the mirror, one needs M
independent measurements (or more) and a mono detector
measurement is not sufficient. It is thus required to add one
dimension (1D) in the measurement process. Let's sweep, for
example, the source wavelength while measuring, as a Network
Analyzer\inxx{network analyser} does, the response to light (i.e
attenuation and phase change) of the mirror DUT (Device Under
Test). Making a frequency to time Fourier transform allows to
determine the time response of the DUT to ultra short pulses.
This sweeping technique is expected to be more sensitive (one
photon sensitivity), more precise (phase sensitive) and must work
with a higher dynamic range (since it detects an amplitude and
not an intensity) than the short laser pulse technique
\cite{pulse}. Solving the 1D diffusion problem is then possible.
Let's consider the reflection configuration. The first reflected
signal allows to determine the refractive index of the first
slab. One can thus calculate the second slab incident field at
any time by accounting of the first slab reflection. Calculation
of the successive slabs index can thus be done by iteration of
this stripping process.

This 1D "gedanken" experiment illustrate the advantage of
Heterodyne Holography in performing a complete measurement of the
field that allows a further powerful data analysis. It also shows
that sweeping the coherent source wavelength is an essential
ingredient to go further. We must notice that in the gel
experiment (Fig.\ref{gel}), the observed noise floor is related
to speckle\inxx{speckle,noise}\inxx{noise,speckle}. Speckle comes
from the multi scattered photons, which remain coherent in time
(because our Helium Neon source have a long coherent length), and
which are detected efficiently by our heterodyne system. OCT and
short laser pulse experiments \cite{pulse} use low time coherent
source to filter off in time the multi scattered photons. We
intend to include time filtering in Herodyne Holography to
suppress most of the diffused photons . This can be done either
by sweeping the source wavelength, as mentioned above, or by
using a OCT low coherent source. With this trick, Heterodyne
Holography, which is already a interesting tool, is expected to
be very powerful in studying diffusing media.

This work was supported by  Thomson-CSF Optronique and funded bt
DGA under contract n°98 10 11A.000.







\end{document}